# What's so Hot about Electrons in Metal Nanoparticles?

Gregory V. Hartland,[a,*] Lucas V. Besteiro,[b] Paul Johns,[a] and Alexander O. Govorov[b,†]

[a] *Department of Chemistry and Biochemistry, University of Notre Dame, Notre Dame, IN 46556-5670*

[b] *Department of Physics and Astronomy, Ohio University, Athens OH 45701*

**Abstract:**

Metal nanoparticles are excellent light absorbers. The absorption processes create highly excited electron-hole pairs and recently there has been interest in harnessing these hot charge carriers for photocatalysis and solar energy conversion applications. The goal of this Perspectives article is to describe the dynamics and energy distribution of the charge carriers produced by photon absorption, and the implications for the photocatalysis mechanism. We will also discuss how spectroscopy can be used to provide insight into the coupling between plasmons and molecular resonances. In particular, the analysis shows that the choice of material and shape of the nanocrystal can play a crucial role in hot electron generation and coupling between plasmons and molecular transitions. The detection and even calculation of many-body hot-electron processes in the plasmonic systems with continuous spectra of electrons and short lifetimes are challenging, but at the same time very interesting from the point of view of both potential applications and fundamental physics. We propose that developing an understanding of these processes will provide a pathway for improving the efficiency of plasmon-induced photocatalysis.

---

[*] e-mail: ghartlan@nd.edu; phone: +1 (574) 631-9320
[†] e-mail: govorov@ohio.edu; phone: +1 (740) 593-9430

The ability of metal nanoparticles (**NPs**) to focus light to small volumes has led to their use in a variety of applications, including as substrates for surface enhanced spectroscopies,[1-8] and as light concentrators for solar energy cells.[9-11] These effects arise from the plasmon resonances of the particles, which are coherent oscillations of the conduction electrons.[12-15] These resonances are termed localized surface plasmon resonances (LSPRs) in the current literature to distinguish them from the propagating surface plasmon polaritons (SPPs) of metal surfaces, wires and plates.[16-18] There has been a tremendous amount of work in designing metal nanostructures to control the LSPR frequency, so that it is now possible to engineer nanostructures that can enhance electromagnetic fields at frequencies from the mid-infrared to the ultraviolet.[19-25]

The light concentrating effects of metal nanostructures are a consequence of the enhanced electromagnetic fields that are generated from the LSPR.[12-15] These external fields are amplified when two or more particles are brought into close proximity, and this amplification is at the heart of surface enhanced Raman spectroscopy (SERS).[1-8] The external fields are also responsible for the strong light scattering effects associated with LSPRs.[2, 13, 26-29] The interaction of metal nanoparticles with light also creates internal fields, which cause absorption.[13-15] Absorption is usually considered to be detrimental to surface enhanced spectroscopy and light concentration applications of nanoparticles, as it leads to heating.[30] However, there are some applications that rely on absorption. One well-known example is the use of metal nanoparticles as localized heat sources for photothermal therapy.[31-34] Another application, which is the focus of this Perspectives article, is harnessing the hot electrons created by light absorption for photocatalysis or solar energy conversion.[35-40]

**Plasmon enhanced photocatalysis.** Several different types of plasmon enhanced photocatalysis/energy conversion systems have been investigated. An early example was gold nanoparticles coupled to TiO$_2$ for water splitting.[41-44] Since then plasmonic nanoparticles have been used to increase the efficiencies of oxidation reactions,[45-49] dissociate small molecules,[50-53] and to generate photocurrents in photovoltaic devices.[54-55] All of these processes involve charge transfer from the metal to nearby semiconductor or molecular states. In general there are two mechanisms that are used to explain electron or hole transfer from excited metal nanoparticles to acceptor states in molecules or semiconductors: (i) a sequential excitation/charge transfer process,[36-38, 41-43, 50-55] or (ii) direct excitation of an interfacial charge transfer transition.[35, 39, 44-48] The quantum yields for charge separation depend on a variety of factors, and a major goal of this article is to review what is known about the relaxation processes in metal nanoparticles and how they affect photocatalysis. In the following we discuss the two mechanisms separately, starting with the sequential mechanism.

In the sequential processes a photon excites the LSPR of the nanoparticle, which subsequently decays to yield an excited electron-hole pair.[14-15, 27, 56-62] The electron-hole pairs are distributed over a range of energies, some of which are high enough to allow electrons to tunnel into vacant states of nearby molecules or semiconductors.[36-38, 57-61] The key factors in determining the quantum yield in this case are the timescale for dephasing of the LSPR, the energy distribution of the excited electron-hole pairs, and the rate of energy relaxation for the electrons/holes compared to the rate of charge transfer across the interface.

**Dynamics of plasmons and hot electrons in plasmonic systems.** The dephasing of the LSPR is extremely fast, to the extent that it cannot be accurately measured using conventional ultrafast measurements.[56] A particularly useful approach for studying LSPR dephasing is to measure the linewidths of single metal nanoparticles.[26-27, 63-68] These measurements where first done by Rayleigh light scattering experiments,[26-27, 63-64] but recently elegant single particle absorption measurements have also been applied to this problem.[65-68] The LSPR linewidth has contributions from several different processes: direct decay of the LSPR into excited electron-hole pairs $\Gamma_b$ (known as Landau damping), radiation damping $\Gamma_{rad}$ (energy loss by scattering a photon), and damping due to electron-surface collisions $\Gamma_{surf}$.[27, 56, 64, 67, 69] The surface term is the one that is responsible for the generation of hot electrons, and will be discussed in detail below.[70] Note that $\Gamma_b$ contains contributions from inter-band as well as intra-band ("Drude model") transitions, i.e. $\Gamma_b = \Gamma_{Drude} + \Gamma_{inter-band}$.[14] $\Gamma_b$ is usually assumed to be the same as the decay rate for electrons in the bulk metal,[56] although recent calculations have shown significant variations in $\Gamma_b$ for small particles with respect to the bulk value, when the electronic spectrum becomes discrete.[71] For not too large particles, $\Gamma_{rad}$ and $\Gamma_{surf}$ are proportional to the volume $V$ and the inverse of the effective path length for electrons in the particle $1/l_{eff}$, respectively.[13, 56, 72-73] Thus, the total linewidth can be written as:

$$\Gamma = \Gamma_b + \Gamma_{rad} + \Gamma_{surf} = \Gamma_b + 2\hbar\kappa V + A\frac{v_F}{l_{eff}} \qquad (1)$$

where $v_F$ is the Fermi velocity, and $\kappa$ and $A$ are constants that characterize the efficiency of radiation damping and electron-surface scattering, respectively.[56, 67] For spheres $l_{eff}$ is simply proportional to the diameter. It is important to note that both the quantum and classical mechanics treatments for electron-surface interactions give the same size scaling.[67, 72-74]

**Figure 1(a)** shows example spectra of single gold nanorods recorded by Rayleigh scattering measurements.[63] The two panels show spectra from samples with different average widths (8 and 14 nm), and the three spectra in each panel are three different particles. The lengths of the nanorods were adjusted so that the two samples had the same aspect ratio.[63-64] The spectra are broader for the narrower rods, which is due to increased electron-surface scattering. Analysis of this data yields estimates for $\kappa$ and $A$.[63] The nanorods in these experiments were coated in surfactant, which means at $\Gamma_{surf}$ has contributions from the adsorbed molecules as well as the metal surface (see below).

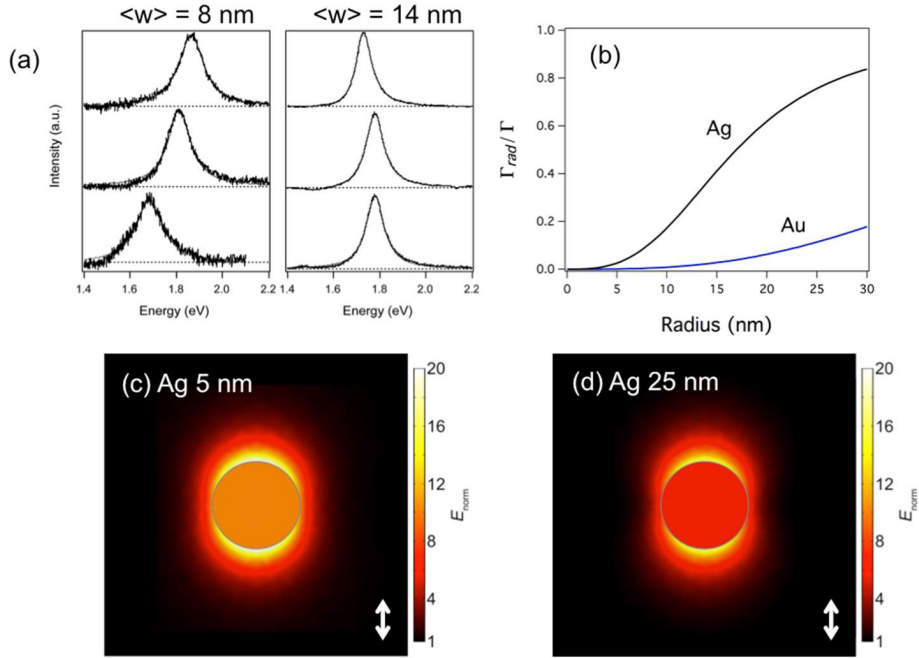

**Figure 1. (a)** Rayleigh scattering spectra for single gold nanorods, reproduced with permission from Ref. [63]. The average width of the rods is given in the figure, and each panel shows spectra from three different nanorods. **(b)** Relative contribution of radiation damping to the plasmon damping $\Gamma_{rad}/\Gamma$, as calculated through $\sigma_{abs}/\sigma_{ex}$. **(c) and (d)** Images of the normalized electric fields for 5 nm and 25 nm radius spherical Ag particles in water. The calculations in (c,d) were performed at the peak of the plasmon resonance for the spherical Ag particles and the images have been scaled so that the particles appear to be the same size. The double-headed arrows show the polarization of the light field.

Measurements for different sized particles show that the LSPR lifetime is ≤10 fs, and that radiation damping dominates over the non-radiative processes for nanoparticles with diameters larger than 20 nm.[56, 67] This is shown in **Figure 1(b)**, where a plot of $\Gamma_{rad}/\Gamma$ is presented in for Ag and Au nanospheres in water. In this figure $\Gamma_{rad}/\Gamma$ was calculated through $\sigma_{scat}/\sigma_{ex}$ where the extinction cross-section $\sigma_{ex}$ is the sum of the absorption and scattering cross-sections ($\sigma_{ex} = \sigma_{abs} + \sigma_{scat}$), and the cross-sections are evaluated at the peak of the plasmon resonance. The correlation between cross-sections and linewidths is possible because both quantities are proportional to the dissipated energy, see Ref. [18]. The cross-sections were calculated through Mie theory, using the dielectric constant data from Ref. [75]. As size increases more incoming photons are scattered and radiation damping becomes more important. Note that $\Gamma_{rad}$ makes a stronger relative contribution to the damping for Ag compared to Au. This is due to the interband transitions of Au, which increase absorption in the region of the plasmon resonance.

**Figure 1(c) and (d)** shows images of the normalized electric fields for 5 nm and 25 nm silver particles in water, again at the peak of the plasmon resonance. The calculations include contributions from electron-surface scattering,[63-64] with an electron-surface scattering parameter of $A = 0.7$ (which is appropriate for Ag[67]). The most noticeable feature is that the fields are strongly enhanced just outside the nanoparticle surface. This effect is what gives rise to the field enhancements in SERS. However, the field inside the particle is also enhanced, and it is this field that causes photon absorption, see below. Note that the field enhancements are similar for the two different sizes. This is due to compensating effects from radiation damping and electron-surface scattering. For the large particles electron-surface scattering is not important, but the LSPR is broadened by radiation damping which reduces the field enhancement. For the small particles the reverse is true - radiation damping is not important but the particles suffer electron-surface scattering.

The fast decay time for the LSPR means that, for all intents and purposes, the hot electron-hole pairs are created instantaneously after photoexcitation. The electron-hole pairs produced by decay of the LSPR will be distributed over a range of energies in the metal's electronic bands.[57-61, 76] For bulk metal surfaces this distribution has been measured through ultrafast photoelectron spectroscopy experiments.[77-81] The results from the measurements show that the electrons initially have a non-thermal distribution that rapidly relaxes to a thermal (Fermi-Dirac) distribution via electron-electron scattering processes.[77-81] Photoelectron spectroscopy measurements are more challenging for nanoparticles. Regular (all optical) ultrafast transient absorption experiments on Ag nanoparticles have been used to measure the timescale for electron-electron scattering for metal nanoparticles with different sizes.[76, 82-83] However, these experiments do not provide direct information about the energy distribution of the hot electron-hole pairs.

Fortunately, insight into the energy distribution of the excited charge carriers can be obtained through theory.[40, 57-62, 70, 84-90] In the steady state (CW) excitation regime there are two typical physical situations. The first case is a steady-state hot electron distribution in isolated NPs. Another physical situation is current injection from a plasmonic nanostructure to an electric contact.[88-90] **Figure 2** shows the calculated distributions of excited carriers in optically-excited nanospheres of various radii.[70] These distributions were computed in a model that includes two electron relaxation times, one for the momentum and another for the energy (see also Fig. S1 in supporting Information).[61, 70] The approach of two relaxation times allows us to describe the two key observations of the plasmonic dynamics.[61, 70] (1) Plasmons exhibit fast dephasing due to the short momentum relaxation time that can be extracted from the plasmonic-peak broadening in the absorption spectra. Within the Drude model, this time is given by the parameter $\Gamma_p = \hbar / \tau_p$ (Eq. S5). For gold and silver plasmons, the dephasing times are obviously short, 8 and 33 fs, respectively (Table S1). (2) The cooling of hot electrons due to photon emission is typically much slower and occurs in the sub-ps range. Such relaxation time can be taken from time-resolved experiments.[56] **Figure 2** shows the non-equilibrium distribution of electrons ($\delta n$) as a function of electron energy in the CW illumination regime. Above the Fermi level ($\varepsilon > E_F$, with $E_F = 5.5\ eV$ in gold) the non-equilibrium population is positive, and below the Fermi level it is negative. Thus, the excited carriers above the Fermi level should be regarded as plasmonic electrons, and empty states below the Fermi level should be considered as holes.

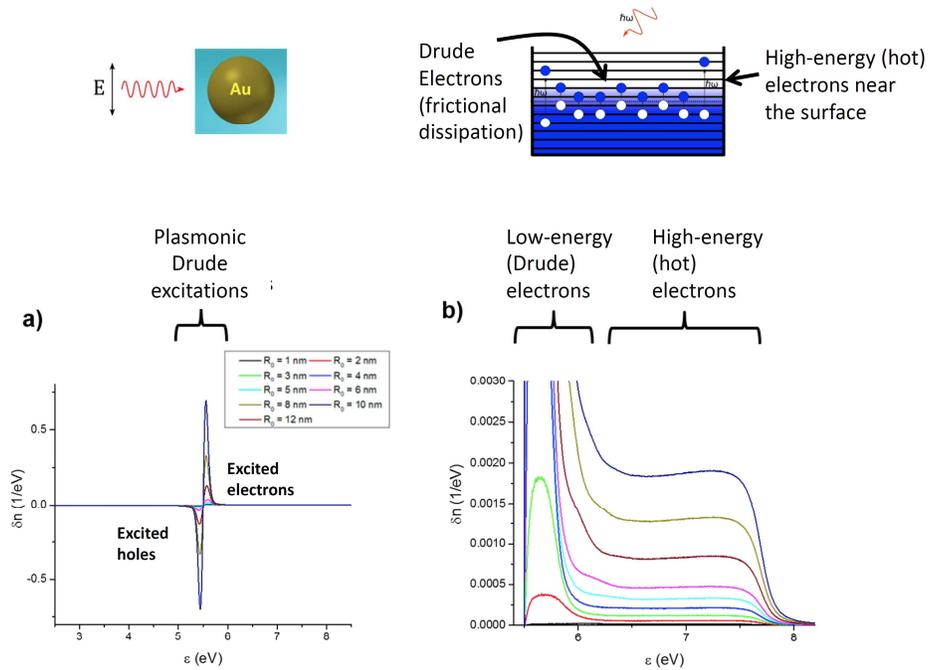

**Figure 2.** (a) Calculated distribution of excited electrons and holes in the Fermi sea of a nanoparticle in the CW illumination regime. (b) Close look at the plateau region with the hot electrons generated by quantum surface-assisted transitions. Upper insets: Plasmonic nanosphere and the Fermi sea with excited electrons and holes. Excited electron-hole pairs in the bulk have small energies, whereas the carriers generated near the surfaces are more energetic. This figure was adopted from Ref. [70].

The striking feature of the distributions in **Figure 2** is the presence of two types of excited carriers. Excited carriers with low excitation energies, near the Fermi level (Drude electrons and holes), form the coherent electron currents, which are described by the classical Drude model. These electrons are responsible for the plasmon oscillation. The high-energy (hot) electrons and holes, which can be used for photocatalysis, occupy the flat plateau regions of the distributions in **Figure 2**. These energetic electrons are created via the quantum optical transitions near the surfaces, and represent the quantum effect of surface scattering. These transitions allow the occupation of high excitation energy states, and become possible due to breaking of linear momentum conservation. The different decay processes are shown schematically in Figure 3(a).

The breaking of linear momentum conservation in a NP is a key mechanism that can also be interpreted in terms of the discretization (or quantitation) of the electronic states in the confined volume of a NP. This quantization involves, of course, the surfaces of the structure. In a large NP, the quantization and the surface scattering effects become equivalent, as it was shown theoretically in Refs. [57, 59, 61, 70]. Another name for the decay of a plasmon through surface scattering, which is used sometimes in the literature, is Landau damping.[71] Although the term "Landau damping" is better suited for the case of a plasmonic 3D running wave, as discussed for classical and quantum plasmas.[91-92]

**Figure 3(b)** shows the rates of generation of the Drude (low energy) and high-energy electrons.[70] The Drude electrons represent the majority of carriers in nanocrystals with relatively large sizes (> 4 nm) and only small nanocrystals have comparable numbers of hot electrons. The reason is that the generation of hot electrons is a surface phenomenon, whereas the Drude electrons represent a bulk effect. However, the energy efficiency of hot-electron production $Eff_{hot-electrons}$ remains relatively large even at sizes of 20 – 30 nm (Figure 3c). The energy efficiency, or the quantum yield for producing hot electrons from the plasmon, reports on the importance of surface scattering:

$$Eff_{hot-electrons} = QP_{plasmon} = \frac{Q_{hot-electrons}}{Q_{tot}}, \qquad (2)$$

where $Q_{hot-electrons}$ and $Q_{tot}$ are the absorption of light due to the generation of hot electrons and the total absorption of a NP (for details see Supporting Information). We note that the efficiency parameter (Eq. 2) also describes the material efficiency of hot-electron production, $Eff_{chem} = Q_{hot-electrons} / N_{metal}$, where $N_{metal}$ is the total number of metal atoms in a solution. This parameter is important for practical photocatalytic applications and it decays with increasing NP size, $Eff_{chem} \propto 1/a_{NP}$ (see also discussion below).

We should note that the rates for low and high excitation energies are very sensitive to the choice of parameters.[57, 70, 84] In particular, the ratio of the rates $Rate_{low-energy,Drude} / Rate_{high-energy} \propto \Gamma_{Drude}$,[61] where $\Gamma_{Drude}$ is the Drude relaxation parameter (see Supporting Information), which describes the rate of relaxation of the electron momentum. This Drude parameter also contributes to the bulk broadening $\Gamma_b$ that appeared in Eq. 1. **Figure 3b** clearly demonstrates the importance of the choice of material system. Silver NPs with a small Drude relaxation rate and sharper plasmonic peak are preferable over gold NPs since the Drude broadening in gold is about four times larger, $\Gamma_{Drude,Au} / \Gamma_{Drude,Ag} \sim 4$ (see Supporting Information). Therefore, silver permits stronger quantum effects and larger hot-electron generation rates.

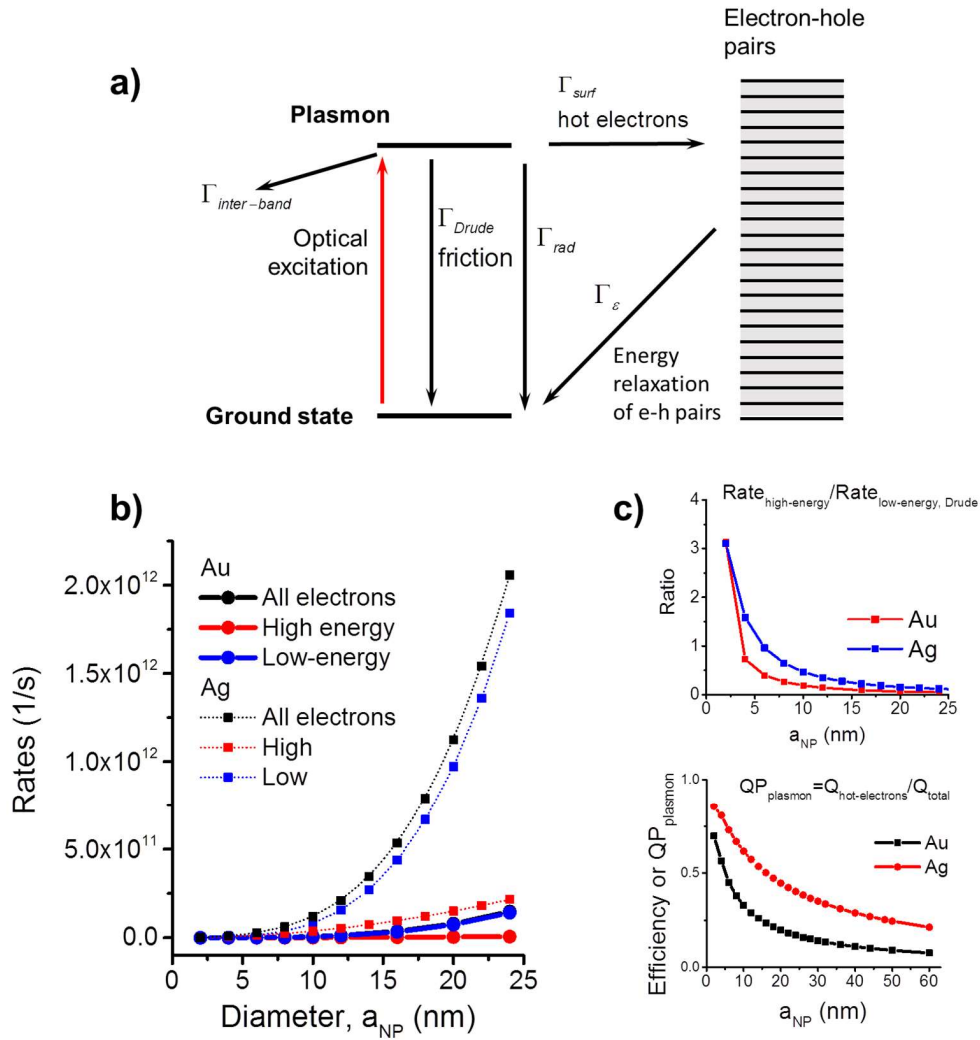

**Figure 3: (a)** Energy diagram for decay of a plasmon in a nanocrystal which includes both classical (Drude-like friction) and quantum (hot electrons) mechanisms. **(b)** Rates of generation of Drude and hot electrons, calculated for Au and Ag nanospheres. **(c)** The upper panel shows the ratio between the rates of generation for high and low energy electrons. In the lower panel: The energy efficiency of hot-electron production. The data are partially taken from Ref. [70]. In this figure, the parameters of excitation are the following: $\hbar\omega = 2.2$ eV (gold), $\hbar\omega = 3.1\,eV$ (silver) and $I_0 = 3.6 \cdot 10^3\,W/cm^2$.

Figure 3(a) shows an energy level diagram for the different decay processes. Note that the surface scattering term is responsible for the creation of hot electrons. A quantum equation for the rate of generation of hot electrons through surface scattering for NPs of an arbitrary shape can be written as an integral over the surface (see Supporting Information for some details):[93]

$$Rate_{high-energy} \approx \frac{2}{\pi^2} \times \frac{e^2 E_F^2}{\hbar} \frac{1}{(\hbar\omega)^3} \int_{S_{NC}} |E_{normal}(\theta,\varphi)|^2 \times ds, \qquad (3)$$

where $E_{normal}$ is the electric field normal to the nanocrystal surface and the integral is taken over the whole surface. The electric field $E_{normal}$ excites hot electron-hole pairs near the boundaries of a NP and it should be taken inside the NP, see Figure 1. The parameter $E_F$ is the Fermi energy. An important quantum factor $(\hbar\omega)^3$ comes from the summation over all quantum optical transitions. For a small nanosphere, the rate of generated hot carriers becomes[93]

$$Rate_{hot-electrons} = \frac{2}{\pi^2} \times \frac{e^2 E_F^2}{\hbar} \frac{1}{(\hbar\omega)^3} \frac{\pi}{3} a_{NP}^2 \left| \frac{3\varepsilon_{matrix}}{2\varepsilon_m + \varepsilon_{metal}} \right| \frac{2\pi}{c_0 \sqrt{\varepsilon_{matrix}}} I_0, \qquad (4)$$

where $I_0$ is the intensity of incident light, $a_{NP}$ is the NP diameter, and $\varepsilon_{matrix}$ and $\varepsilon_{metal}$ are the dielectric constants of the matrix and metal, respectively. Because the rates for generating high-energy and low-energy electrons scale approximately as the surface area and volume, respectively, the number of high-energy electrons in a large nanocrystal is small in proportion to the number of low-energy electrons (see **Figures 2** and **3(b)**). The ratio between their rates of generation can be approximated by:[57, 59, 70]

$$\frac{Rate_{high-energy}}{Rate_{low-energy,Drude}} = const. \frac{l_{mfp}}{a_{NP}} \frac{v_F/a_{NP}}{\omega}, \qquad (5)$$

where $l_{mfp}$ is the electronic mean free path, and $v_F$ is the Fermi velocity of the metal. To summarize the consequences coming from Eqs. (2), (3) and (5), we observe an important property of plasmonic NPs - a different size-dependence for the ratios of rates and energy absorptions:

$$\frac{Rate_{high-energy}}{Rate_{low-energy,Drude}} \propto \frac{1}{a_{NP}^2},$$

$$Eff_{hot-electrons} = \frac{Q_{hot-electrons}}{Q_{tot}} \propto \frac{1}{a_{NP}}.$$

Now we will make more observations. As one can see from **Figures 3(b,c)**, the ratio $Rate_{high-energy}/Rate_{low-energy,Drude} \sim 1$ for a 2 nm gold NP, but it decreases rapidly for larger sizes. The decay rate of the plasmon for the hot-electron pathway and the rate of generation of hot electrons in a nanocrystal are, of course, related. Specifically, the simple calculation in Ref. [93] and also simple physical arguments show that the quantum dissipation and the high-energy rate are proportional:

$$Q_{hot-electrons} = \hbar\omega \cdot Rate_{high-energy}.$$

Therefore, the surface decay rate given by Eq. 1 can be also written as

$$\Gamma_{surf} = \frac{\hbar\omega \cdot Rate_{high-energy}}{E_{plasmon}} = A\frac{v_F}{l_{eff}} \quad (6)$$

where $E_{plasmon}$ is the energy stored in the plasmon (this energy has both kinetic and potential components) and $l_{eff} \approx a_{NP}$, the effective size of the NP. Equation (6) shows that hot electron generation can be probed by investigating surface scattering. However, for the majority of NPs, the observation of surface-scattering induced broadening of the plasmon is not easy, since the NPs are typically of relatively large sizes. The rate $\Gamma_{surf}$ can only be clearly identified from the plasmon peak broadenings for small NP sizes.[94-95] At the same time, photochemistry and photocurrent experiments, which are sensitive to the excitation of high-energy electrons near the NP surfaces, provide evidence that high-energy electrons have been created in large nanoparticles.[35, 37, 39, 45, 96-104]

Although the phenomenological picture of plasmon decay in Figure 3a is very convenient to analyze and describe photocatalytic, time-resolved and photo-current experiments, there is another kinetic representation that is often used, especially in time-resolved studies (see a review Ref. [67]). **Figure 4** illustrates it. Electrons are excited from the continuum of occupied states via two types of transitions, low energy (frictional) and high energy (quantum) (Figure 4a). In a fs-pulse experiment, an initial distribution of excited carriers contains some number of high-energy electrons that relax through e-e and e-phonon pathways. Current literature typically depicts short-lived hot-electrons as two flat regions (Figure 4b), while the many-body theory[61] should produce somewhat different transient distributions (Figure 4c). Since the frictional transitions are very active (strong Drude currents), the distribution of excited electrons during the fs-pulse should have a large number of low-energy electrons and some number of hot electrons, as shown in Figure 4c. The details and differences are still under discussion in the literature and should be debated further.

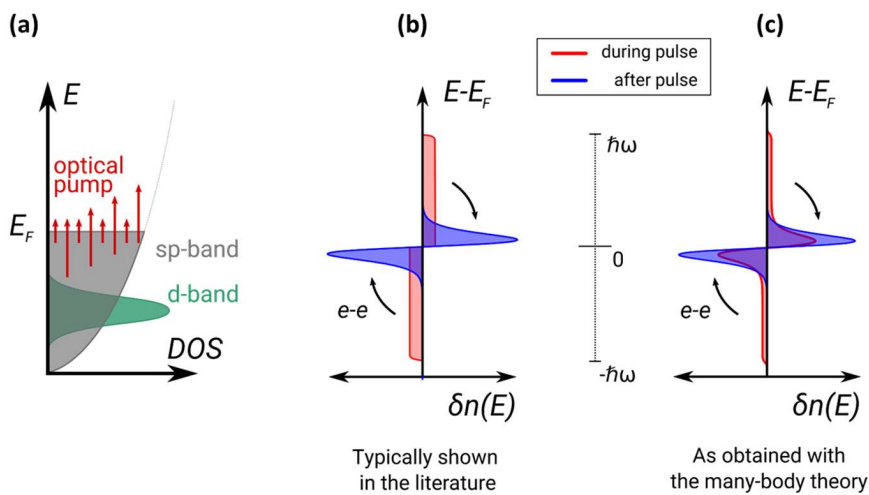

**Figure 4**: Another representation of plasmonic excitation and time dynamics involving the density of states (DOS) (a) and transient hot-electron populations (b,c). The qualitative picture (b) is commonly seen in current literature, while the picture (c) is based on the analysis of quantum kinetics involving the density-matrix formalism.[61] In the picture (c), the distribution of hot electrons has low-energy Drude electrons already during the fs-excitation pulse.

The above analysis shows that highly excited electrons, which are a prerequisite for photocatalysis by the sequential mechanism, are only produced in large proportions for small particle sizes. Whether the hot electrons created by light absorption can participate in photocatalysis/energy conversion depends on their relaxation times compared to the rate of interfacial charge transfer.[81] We first consider the relaxation timescales. The initial non-thermal electron distribution created by dephasing of the LSPR relaxes by electron-electron scattering. The best estimates for the electron-electron scattering times in small metal nanoparticles are ca. 200 fs.[76, 82] Electron-electron scattering creates a thermal distribution of electrons at an elevated electronic temperature, which subsequently relaxes on a few ps timescale through electron-phonon coupling.[56, 105] There are significantly less highly excited electrons in the thermal electron distribution compared to the initial non-thermal distribution. Using the electronic heat capacity for bulk gold, we estimate that a single visible (530 nm) photon will induce a 1370 K increase in the electronic temperature for a 2 nm diameter particle. Even though this is a large temperature increase, at this temperature only 4% of the excited electrons will have an energy > 0.5 eV above the Fermi level, the majority of the excited electrons will be in states around the Fermi level, as in **Figure 2**. Thus, any hot electron processes must occur before electron thermalization.

The timescales for interfacial electron transfer at metal surfaces are less well established compared to the internal electron relaxation times. For metal nanoparticles coupled to semiconductors, ultrafast measurements have implied very fast (< 100 fs) metal-to-semiconductor electron transfer times.[42, 96, 106] Thus, electron transfer into acceptor states of the semiconductor are possible from both the initial non-thermal electron distribution and the thermal distribution created by electron-electron scattering.[35-38, 41-44, 54-55, 96, 106] We would expect the major contribution to come from the non-thermal electron distribution, despite the shorter lifetime, because of the larger number of highly excited electrons.

The situation is more complicated for electron transfer to molecular states.[107] Molecule to semiconductor electron transfer reactions have been shown to occur on ultrafast (sub-ps) timescales,[108] but in this case the fast rate arises from the high density of states of the acceptor (the solid).[109-110] Indeed, the reverse semiconductor to molecule charge transfer reaction can be quite slow.[111-114] Thus, it seems unlikely that the sequential mechanism could cause a significant amount of charge-transfer to molecular states. The difficulty in reconciling the ultrafast relaxation times for the electrons in the metal nanoparticles with the slow (> 10-100 ps) interfacial electron transfer times to molecular states has led to the development of a fundamentally different explanation for plasmon enhanced molecular photocatalysis: excitation of an interfacial change transfer transition.[39, 47-48, 96, 115] In this mechanism a charge transfer transition is directly excited, so that the internal relaxation of electrons within the metal nanoparticle is irrelevant. The direct excitation mechanism has been used to explain plasmon induced oxidation reactions,[39] as well as the high quantum yields observed for photo-induced electron transfer from Au nanoparticles to attached semiconductor quantum dots.[96]

In the direct mechanism, small molecule plasmon enhanced photocatalysis reactions proceed through a "dissociation induced by electronic transitions" (DIET) process.[116-122] In DIET excitation of the charge transfer transition transiently populates a surface bound anion state of the molecule. In general this state will be vibrationally excited, and will rapidly relax by vibrational cooling followed by electron transfer back to the metal.[39] However, when the excitation rate exceeds the relaxation rate the bond can become activated, leading to dissociation. This creates reactive species that subsequently participate in further chemical reactions. DIET has been extensively studied for metal surfaces, where high light fluxes are typically needed for reaction.[116-120] For nanoparticles it is proposed that the high fields created in "hot spots" (junctions between two or several nanoparticles) enable the DIET process.[115] Evidence for the DIET mechanism comes from the intensity and temperature dependence of the reaction rates.[46] Surface enhanced Raman measurements have also been used to explore how DIET like processes can funnel energy into molecular vibrations.[115, 123-124]

One of the fundamental differences between the direct and sequential mechanisms discussed above is that in the sequential mechanism plasmon dephasing occurs before electron transfer,[36-38] whereas in the direct mechanism the presence of charge transfer transitions leads to plasmon dephasing.[96] This means that, in principle, the direct mechanism should produce an extra contribution to the LSPR linewidth. The fact that adsorbed molecules can increase the LSPR linewidth through excitation of charge transfer type transitions was recognized in the 1970s.[125-130] This processes is known as "Chemical Interface Damping" (CID), and was studied through ensemble measurements before the development of single particle spectroscopy.[125-129] In absorption measurements of single particles the large variations in the LSPR linewidth for surfactant-coated particles has been attributed to CID,[131-133] which has led to the use of silica-coated particles for precise linewidth measurements.[95, 134] Indeed, recent single particle scattering measurements have shown that carefully removing the surfactant layer around gold nanorods creates very narrow linewdiths.[135]

The potential importance of CID in plasmon-induced photocatalysis has renewed interest in studying this effect from both experimentalists and theoreticians. For example, theoreticians have begun to use electronic structure calculations to examine how molecules interact with the electrons in metal nanoparticles.[136-137] In an elegant set of experiments, Link and co-workers used the change in linewidth for gold nanorods with and without a layer of graphene to estimate the timescale for electron transfer from gold to graphene.[138] Foerster and co-workers also recently showed that the CID contribution to the LSPR linewidth scales with the particle's dimensions in the same way as electron-surface scattering.[139] This means that CID can be included in the expression for the LSPR linewidth (Equation (1)) by simply writing $A = A_{surf} + A_{CID}$, where $A_{surf}$ represents the effect of the nanoparticle surface and $A_{CID}$ is the effect from adsorbed molecules.[139] This analysis predicts an increase in linewidth with surface adsorption (molecular species at the surface of the particle introduce additional decay channels for the plasmon that will increase damping). However, CID experiments that involve ligand exchange could show a decrease in linewidth depending on the exact nature of the molecules being exchanged. Clearly it will be interesting to characterize how $A_{CID}$ varies with different types of molecules, the degree of surface coverage and the LSPR frequency.

**The use of hot spots and special designs to produce hot electrons.** Plasmonic hot spots are small regions of space inside and around metal nanostructures where the electromagnetic fields become strongly amplified. Such spots appear in narrow gaps between nanocrystals or at tips and apexes of nanocrystals with complex shapes, like nanocubes or nanostars (**Figure 5**). It is now well established that hot spots play an important role in SERS.[2-3] Such small volumes with strong electromagnetic fields can also contribute to and even dominate hot electron photochemistry.[46, 101] The process of hot electron generation due to hot spots involves two components: (1) Enhancement of the magnitude of the electric field in the hot spot and (2) breaking of linear momentum of the electron in the hot spot due to a strongly non-uniform field. Theoretical calculations have shown that both mechanisms can contribute to the enhanced generation of energetic carriers (**Figure 5**).[59, 70, 93] In particular, it was found that the aforementioned classical mechanism (1) is not sufficient to understand the calculated quantum generation of hot electrons in nanocubes[59] and plasmonic dimers.[70] Another important manifestation of the hot spot generation of carriers was reported recently in a time-resolved experiment on a meta-structure with extended and strong hot spots.[98] Extended hot spots can be achieved, for example, in planar super-absorber meta-structures with a metallic reflecting layer.[98] In contrast to the typical ps dynamics of an electron system in plasmonic nanocrystals,[56, 76] the paper by Harutyunyan et al.[98] reported anomalous relaxation kinetics in which an ultra-fast fs component dominated. This behavior was explained by the ultra-fast electron-electron scattering of energetic carriers.

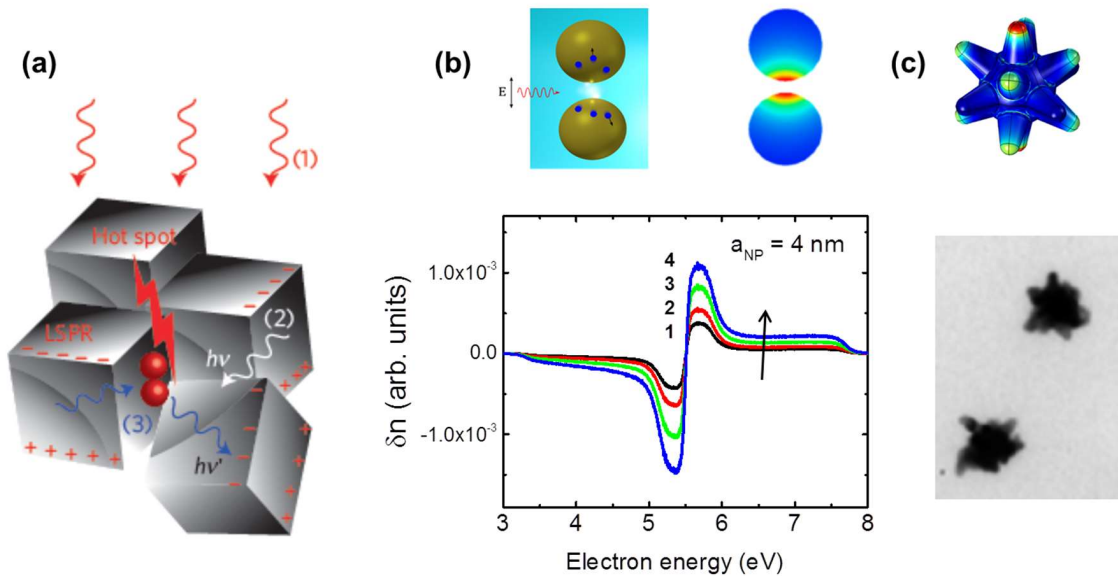

**Figure 5**: **(a, b)** Illustrations of photo-chemistry experiments, from Ref. [46] Panel (a) shows a cartoon of hot electron generation in an aggregate of nanocubes with a hot spot. (b) Calculation of hot electron generation enhanced by the hot spot in a gold NP dimer; the curves show the hot-plasmon distributions and are shown for different NP-NP gaps, $\Delta = \infty, a_{NP}/2, a_{NP}/4, a_{NP}/6$. One can see how the hot-electron plateaus strongly grow for small gaps. Insets: Model of a plasmonic dimer with a hot spot near the gap and the electromagnetic field map showing strong enhancement in the hot-spot region. Adapted from Ref. [70]. (c) Model of a plasmonic nanostar with color showing its calculated surface fields, adapted from Ref. [93]. Now the hot-spot effect occur near the tips. The lower panel is an image of colloidal nanostars taken from the experimental study of Ref. [101].

**Hybrid exciton-plasmon systems.** For the Au nanoparticle-semiconductor quantum dot experiments performed by Lian and co-workers, the LSPR of the gold particles is broadened beyond recognition.[96] This suggests very strong coupling, and raises an interesting point about our understanding of CID. The view of CID as an additional broadening mechanism for the LSPR is essentially a perturbation theory/weak coupling description: the LSPR is modified by CID but maintains its identity. However, it is also possible to create strongly coupled, hybrid plasmon-exciton states.[140-142] These states have been extensively studied for the propagating surface plasmon polaritons (SPPs) of metal surfaces.[142-154] Dispersion curves (plots of the SPP frequency versus wavevector) of thin metal films coated with j-aggregates or semiconductor quantum dots can show avoided crossings,[142-154] which provide a direct way of measuring the coupling between the SPPs and the localized exciton transitions associated with the j-aggregates or quantum dots. Strongly coupled plasmon-exciton states have also been identified for the LPSRs of particles, but here the spectral signatures of coupling are more subtle.[155-169] Information about the coupling strength can be obtained from splittings in the extinction[155-162] or scattering spectra of the LSPR, [163-167] or in emission spectra from the adsorbed molecules. [168-169]

An example of an avoided crossing for CdSe quantum dots coupled to SPPs on a Ag surface is shown in **Figure 6**. At the avoided crossing the two states are equal mixtures of the SPP and exciton wavefunctions:[140, 170]

$$|\psi^+\rangle = (|SPP\rangle + |exciton\rangle)/\sqrt{2} \qquad (7)$$

$$|\psi^-\rangle = (|SPP\rangle - |exciton\rangle)/\sqrt{2}$$

The properties of the mixed plasmon/exciton states are still under active investigation. For example, there have only been a handful of dynamics measurements, and at present it is not clear how the lifetimes of the coupled states are related to the initial SPP/exciton states.[147, 150, 170-172] Another interesting problem regarding the exciton-plasmon interaction is the exciton-plasmon coupling in the quantum regime of a single exciton, where the absorption line-shape can become Fano-like.[173-175] This quantum regime was only recently observed and reported in Ref. [176].

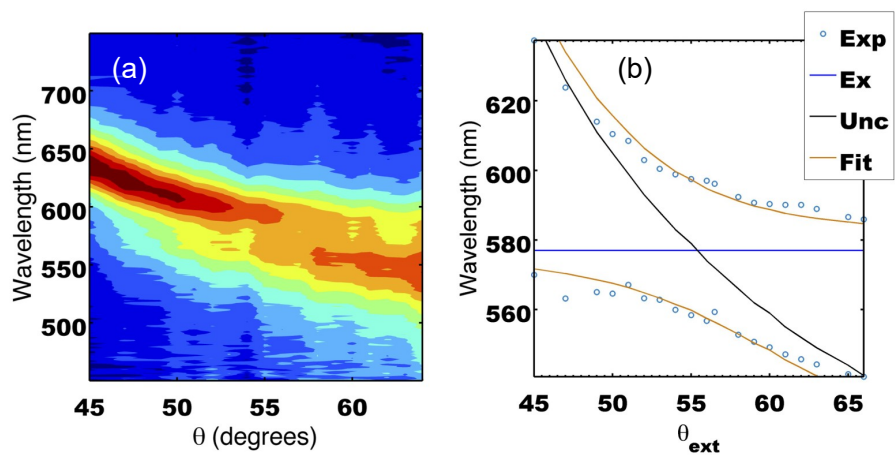

**Figure 6:** Strong coupling between plasmons and excitons for CdSe quantum dots coupled to a thin Ag film. The angle is proportional to the wavevector for the SPP of the Ag film. **(a)** Experimental spectra versus angle, **(b)** peak wavelength extracted from the spectra versus angle data. Reproduced with permission from Ref. [170].

The coupling in the hybrid plasmon-exciton states is typically described using the same language as that for molecules in optical cavities, that is, as arising from the interaction between the transition dipole of the exciton and the electromagnetic field associated with the plasmon.[140-141] On the other hand, in the CID mechanism the coupling is between the wavefunctions for the LSPR and the molecular/semiconductor acceptor states. Reconciling these different approaches, and developing a rigorous wavefunction level theory for the coupling between plasmons and excitons that spans strong and weak coupling will clearly be a challenge.[57, 177] However, improving our understanding of CID could enable the development of efficient plasmon-enhanced molecular photocatalysis reactions.

The above discussion highlights the connection between the spectroscopy and dynamics of metal nanoparticles, and their applications in plasmon induced photocatalysis and solar energy conversion. Considerations of the rates of interfacial charge transfer reactions and the relaxation times for electrons in metal nanoparticles implies that efficient photocatalysis in molecular systems is most likely to occur through direct excitation of interfacial charge transfer transitions. These transitions can be studied by observing how the LSPR linewidth depends on the presence of surface bound molecules,[138-139] and we believe that it will be interesting to connect linewidth measurements with measurements of plasmon induced photocatalysis. For electron transfer to semiconductors the sequential mechanism is also feasible. In this case linewidth measurements may also be able to tease out the relative contributions from the direct and sequential charge transfer processes. On the theory side, more work is needed to develop a wavefunction level description of the interfacial charge transfer states.[57, 177] This is a difficult task for several reasons: first metal nanoparticles are very large systems with many degrees of freedom. Second, such a theory should be able to span the weak coupling to strong coupling regimes, and should be able to generate rate constant information. Third, although researchers achieved a decent understanding of the theory on the generation of hot electrons in confined nanocrystals, direct experimental measurements of the spectral distributions for the classical Drude-like electrons and for the quantum intra-band hot carriers are still needed. Such measurements would also reveal the full potential of usage of plasmonics for photo-chemistry and photo-currents. The unsolved fundamental scientific questions, and the connections to photocatalysis and solar energy conversion, make this an attractive area to work for spectroscopists, theoreticians, and scientists and engineers interested in developing plasmon-enhanced devices.

**Acknowledgements:** G.H. and P. J. acknowledge the support of the National Science Foundation (CHE-1502848) and the Office of Naval Research (Award No.: N00014-12-1-1030). A.O.G. and L.V.B. acknowledge support by Volkswagen Foundation (Germany) and by the Army Office of Research (MURI Grant W911NF-12-1-0407).

**Supporting Information:** The Supporting Information for this paper includes useful equations and parameters for calculating the hot-electron energy efficiencies for Ag and Au.

# Supporting Information

# What's so Hot about Electrons in Metal Nanoparticles?

Gregory V. Hartland,[a,‡] Lucas V. Besteiro,[b] Paul Johns,[a] and Alexander O. Govorov[b,§]

[a]Department of Chemistry and Biochemistry, University of Notre Dame, Notre Dame, IN 46556-5670

[b]Department of Physics and Astronomy, Ohio University, Athens OH 45701

**Quantum efficiency of hot-electron generation**. This short supporting information will provide some useful equations and parameters. The efficiency of hot-electron production is computed in the following way:

$$Eff_{hot-electrons} = QP_{plasmon} = \frac{Q_{hot-electrons}}{Q_{tot}}, \quad (S1)$$

where the total absorption should be calculated as a sum of two terms:

$$Q_{tot} = Q_{classical} + Q_{hot-electrons}. \quad (S2)$$

The first term is the classical absorption in a NP:

$$Q_{classical} = I_0 \sigma_{abs,NP},$$

$$\sigma_{abs,NP} = \frac{1}{c_0 \sqrt{\varepsilon_{matrix}}} \omega \cdot \frac{4\pi R_0^3}{3} \left| \frac{3\varepsilon_{matrix}}{\varepsilon_{metal} + 2\varepsilon_{matrix}} \right|^2 \text{Im}(\varepsilon_{metal}) \quad (S3)$$

---

‡ e-mail: ghartlan@nd.edu
§ e-mail: govorov@ohio.edu

$\varepsilon_{metal}$ and $\varepsilon_{matrix}$ are the dielectric functions of the metal and the matrix, respectively. This term includes the Drude intra-band dissipation and the interband transitions in a bulk metal. The second term in Eq. S2 is of a quantum nature and it should be calculated through the known hot-electron distribution,

$$Q_{hot-electrons} = \frac{1}{\tau_\varepsilon} \int_{|\varepsilon - E_F| > \delta E} d\varepsilon \cdot (\varepsilon - E_F) \cdot \delta n(\varepsilon). \tag{S4}$$

Here $\delta n(\varepsilon)$ is the distribution of non-equilibrium electrons in a NP under CW illumination and $\tau_\varepsilon$ is the energy relaxation time; $\delta E \sim 0.2 eV$ for large sizes and this parameter denotes the threshold energy for hot electrons. Important details for the derivation of the above equations can be found in Ref. S1. The typical structure of the function $\delta n(\varepsilon)$ in NPs in the steady-state regime under continuous illumination is shown in Fig. 2 in the main text.

**On the structure of the dielectric constant of a typical metal and the relaxation rates**. Regarding the structure of the local dielectric constants of noble metals, one can read the textbook [S2]. The dielectric constant of bulk metals, like Au and Ag, has the following structure:

$$\varepsilon_{metal}(\omega) = \varepsilon_b + \Delta\varepsilon_{inter-band}(\omega) + \Delta\varepsilon_{Drude}(\omega), \quad \Delta\varepsilon_{Drude}(\omega) = -\frac{\omega_p^2}{\omega(\omega + i\Gamma_{Drude})}, \tag{S5}$$

where $\Delta\varepsilon_{inter-band}(\omega)$ is the inter-band term, $\varepsilon_b$ is the screening constant due to bound-charges in a metal, $\omega_p$ is the bulk plasmon frequency and $\Gamma_{Drude}$ is the Drude relaxation constant. This important constant is responsible for the description of the friction-like dissipation in an optically-excited metal NP where the induced dynamic electric current creates heat. In the theory used in this paper, we assumed that the Drude parameter is identical to the momentum relaxation rate, i.e.

$$\Gamma_p = \Gamma_{Drude}.$$

This simple conclusion follows from the kinetic consideration of the decay time of electric current in electron plasma [S3]. Below we give the parameters for the Drude dielectric functions, as well as the Fermi energies and the relaxation times for Au and Ag. [S3].

For the traditional model metals, Au and Ag, the detailed dielectric function (Eq. S5) is reduced to the classical Drude function in the wavelength interval where the inter-band transitions become inactive:

$$\varepsilon_{metal,Drude}(\omega) = \varepsilon_{b,Drude} - \frac{\omega_p^2}{\omega \cdot (\omega + i\Gamma_{Drude})}. \qquad (S6)$$

This function is very convenient for a description of plasmonic effects in the IR interval.

Table S1: Parameters used in the calculations of the hot electron distributions.

| Parameter | Au | Ag |
| --- | --- | --- |
| $\Gamma_p = \hbar/\tau_p$ | 0.078 eV | 0.020 eV |
| $\hbar/\tau_\varepsilon$ | 0.0013 eV (0.5 ps) | 0.0013 eV |
| $\hbar\omega_{p,Drude}$ | 9.1 eV | 9.3 eV |
| Work Function | 4.6 eV | 4.7 eV |
| Fermi Energy | 5.5 eV | 5.76 eV |
| $\varepsilon_b$ | 9.07 | 7.0 |

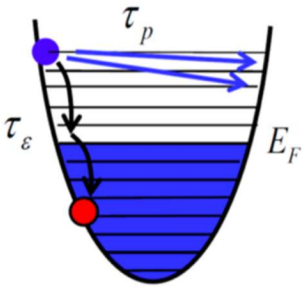

**Figure S1**: Illustration of the single-particle relaxation processes involved in the Kinetic DFT model used in [S3].

**Derivation of the hot-electron rate as a surface integral.** Since this is a perspective article long derivations have no place in it, and we present here only a brief one to complement the main text. This derivation follows a recent paper [S4]. In the kinetic master-equation approach, the generation of high-energy electrons owing to the surfaces of NPs under CW illumination should be written as

$$Rate_{high-energy} = \frac{1}{\tau_\varepsilon} \int_{|\varepsilon - E_F| > \delta E} d\varepsilon \cdot \delta n(\varepsilon),$$

where the integral over energy is taken outside the energy region of the Drude electrons (see Figure S2). For large NPs, the parameter $\delta E$ is given by the thermal energy, $\sim 8 \cdot k_B T$, and in small ones it is given by the single-particle excitation energy, $\sim \hbar v_F (\pi / a_{NP})$ [S1,S2,S4]. We see that the function $\delta n(\varepsilon)$ is essentially flat in the interval of integration (Figure 2 in the main text and Figure S2 below). Then, the function $\delta n(\varepsilon)$ for a smooth surface with an infinite potential wall has the analytical form [S4, S5] for the electron and hole flat regions:

$$\delta n(\varepsilon) = \pm \frac{2}{\pi^2} \cdot \tau_\varepsilon \frac{e^2 E_F^2}{\hbar} \frac{1}{(\hbar\omega)^4} \times \int_{S_{NC}} |E_{normal}(\theta,\varphi)|^2 \times ds,$$

$$E_F + \delta E < \varepsilon < E_F + \hbar\omega$$
$$E_F - \hbar\omega < \varepsilon < E_F - \delta E$$
(S7)

Eq. S7 is essentially an integral of the local hot-electron production over the whole surface of a NP. Therefore, the above equations immediately yield

$$Rate_{high-energy} = \frac{2}{\pi^2} \times \frac{e^2 E_F^2}{\hbar} \frac{(\hbar\omega) - \delta E}{(\hbar\omega)^4} \int_{S_{NC}} |E_{normal}(\theta,\varphi)|^2 \times ds$$

Since the optical energy typically exceeds the thermal energy, i.e. $\hbar\omega \gg \delta E$, we can simplify the above equation (neglecting the term $\delta E$) and arrive to the final equation

$$Rate_{high-energy} \approx \frac{2}{\pi^2} \times \frac{e^2 E_F^2}{\hbar} \frac{1}{(\hbar\omega)^3} \int_{S_{NC}} |E_{normal}(\theta,\varphi)|^2 \times ds. \quad (S8),$$

which is the one given in the main text.

Finally, using the integral (S4) and the energy distribution (S7), we can also obtain an equation for the quantum dissipation:

$$Q_{hot-electrons} = \frac{2}{\pi^2} \times \frac{e^2 E_F^2}{\hbar} \frac{(\hbar\omega)^2 - \delta E^2}{(\hbar\omega)^4} \int_{S_{NC}} |E_{normal}(\theta,\varphi)|^2 \times ds \approx$$
$$\approx \frac{2}{\pi^2} \times \frac{e^2 E_F^2}{\hbar} \frac{1}{(\hbar\omega)^2} \int_{S_{NC}} |E_{normal}(\theta,\varphi)|^2 \times ds \quad (S9)$$

Then, we compare (S8) and (S9) to obtain one useful relation given in the main text (see the equation before Eq. 6):

$$Q_{hot-electrons} = \hbar\omega \cdot Rate_{high-energy}.$$

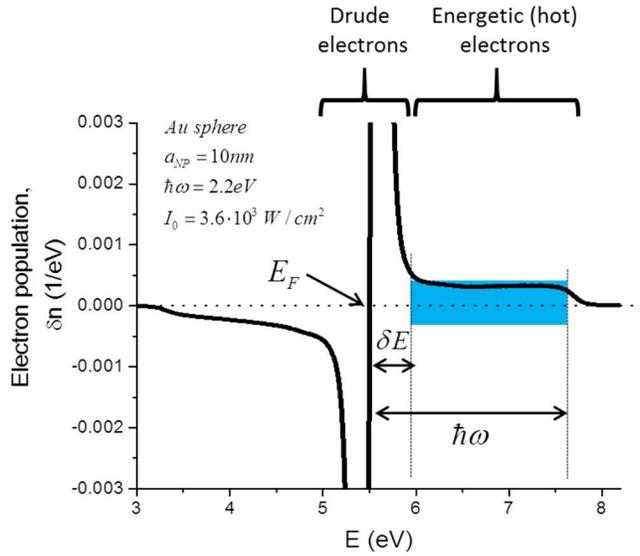

**Figure S2:** Typical distribution of non-equilibrium plasmonic elections under CW illumination in a NP. The blue area shows the interval of integration for the hot electrons.

41